\documentclass[aps,prd,amsfonts,amsmath,showpacs]{revtex4}
\usepackage[dvips]{graphicx}
\usepackage{bm}

\begin{document}

\title{Primordial non-Gaussianity estimator: the inhomogeneous noise effect}

\author{B.Yu}

\email{yubo@chenwang.nju.edu.cn}

\affiliation{Deparment of Physics, Nanjing University, Nanjing 210093, China}

\author{T.Lu}
\affiliation{Purple Mountain Observatory, Chinese Acadamy of Sciences,
Nanjing 210008, China}
\affiliation{Joint Center for Particle, Nuclear Physics and Cosmology, Nanjing University-Purple Mountain Observatory, Nanjing 210093, China}
\date{\today.}
\begin{abstract}
Since the inhomogeneous instrument noise can produce extra non-Gaussianity in the CMB anisotropy, its effect should be carefully subtracted in the primordial non-Gaussianity estimation. We calculate the probability distribution function of the CMB anisotropy for local type of non-Gaussianity, from which the optimal estimator in the general case (inhomogeneous noise and cut sky) is obtained. The new estimator obtained here is different from the popular one, since the inhomogeneous noise and cut sky effects are completely accounted. The CMB anisotropy in the new estimator is noise weighted. The noise weight is different from that used by WMAP Group in their 5-year data analysis. Although it is still difficult to calculate the new estimator rigorously, for the case of the slightly inhomogeneous noise, there exists a series expansion method to compute the new estimator. Each order in the series is suppressed by two factors, $(\frac{\sigma^2}{\sigma^2_{i}}-1)$ and $\frac{C_l}{C^{tot}_{l}}$, which make the method feasible. Through the Edgeworth expansion we can generalize our discussion to other types of non-Gaussianity.
\end{abstract}
\maketitle
\section{Introduction}
The standard inflation predicts a flat universe with scale invariant and gaussian random primordial fluctuation, which is confirmed by the Cosmic Microwave Background (CMB) observation \cite{Spergel:2003cb}. Despite its extreme success in modern cosmology, there are too many inflation models which all give the above predictions. However, with increasing data, it is possible to distinguish those models by some subtle features, for example, the primordial gravitational waves, the running power spectrum, the isocurvature perturbation and the primordial non-Gaussianity \cite{Babich:2004gb}. Among them the primordial non-Gaussianity is an important one.

The primordial non-Gaussianity can be quantified by the 3-point correlation function of curvature perturbation $\Phi (\vec{k})$ (equivalently by the bispectrum $B(k_1,k_2,k_3)$),
\begin{eqnarray}
< \Phi (\vec{k}_1) \Phi (\vec{k}_2) \Phi (\vec{k}_3)> = (2\pi)^3 \delta^3(\vec{k}_1+\vec{k}_2+\vec{k}_3) B(k_1,k_2,k_3).
\end{eqnarray}
Acoording to \cite{Babich:2004gb} the primordial non-Gaussianity can be roughly divided into two types: local type and equilateral type. ``local type'' means $B(k_1,k_2,k_3)$ is large when one of the momenta is small compared with the other two, while ``equilateral'' type means $B(k_1,k_2,k_3)$ is large when the three momenta are of the same order. In the real space the local type of non-Gaussianity can be represented as:
\begin{eqnarray}
\Phi (\vec{x}) = \Phi_L(\vec{x})-f_{NL}(\Phi^2_L(\vec{x})-<\Phi^2_L(\vec{x})>),
\label{phi}
\end{eqnarray}
where $\Phi_L$ is a gaussian random curvature perturbation and $f_{NL}$ describes the strength of the primordial non-Gaussianity.

The standard inflation model predicts only a small non-Gassianity \cite{Maldacena:2002vr}, which is beyond the experimental limit. However, this is changed in multiple-field models, which is investigated in \cite{Seery:2005gb,Vernizzi:2006ve,Lyth:2002my,Sasaki:2006kq} and many other papers. Besides this, non-local inflations and ghost inflation theories can also produce large non-Gaussianity \cite{Barnaby:2007yb,ArkaniHamed:2003uz,Cheung:2007st}. Even in the single field inflation, it is possible to generate large non-Gaussianity \cite{Chen:2006xjb,Chen:2008wn}.

So the non-Gaussianity measurement is important to constrain the inflation dynamics. The general properties of primordial non-Gaussianity in the bispectrum measurement are discussed in \cite{Komatsu:2001rj}. However, the direct measurement of the bispectrum is very time-consuming for WMAP and future data. A fast local type of non-Gaussianity estimator is proposed in \cite{Komatsu:2003iq} and it was applied to the 1-year WMAP data to constrain the non-Gaussianity in \cite{Komatsu:2003fd}. The fast estimator proposed in \cite{Komatsu:2003iq} is originally constructed on Wiener-filtered maps. Later on, \cite{Babich:2005en} proved that it is optimal in weak non-Gaussianity and full sky case. But it is already realized in \cite{Komatsu:2003iq} that the inhomogeneous noise and cut sky play an important role in the non-Gaussianity estimation, since from them extra non-Gaussianity can arise. In \cite{Creminelli:2005hu} a linear term is introduced into the estimator to correct this inhomogeneous noise and cut sky effects. The complete estimator can be written as: 
\begin{eqnarray}
\epsilon_{lin}(a) &=& \frac{1}{S} \sum_{(l,m)} [(< a_{l_1 m_1} a_{l_2 m_2} a_{l_3 m_3} >_{f_{NL}=1}) C^{-1}_{(a)l_1 m_1,l_4 m_4} C^{-1}_{(a)l_2 m_2,l_5 m_5} C^{-1}_{(a)l_3 m_3,l_6 m_6} a_{l_4 m_4} a_{l_5 m_5} a_{l_6 m_6} \nonumber\\
&& -3 (< a_{l_1 m_1} a_{l_2 m_2} a_{l_3 m_3} >_{f_{NL}=1}) C^{-1}_{(a)l_1 m_1,l_2 m_2} C^{-1}_{(a)l_3 m_3,l_4 m_4} a_{l_4 m_4}],
\label{ea}
\end{eqnarray}
where $C_{(a)l_1 m_1,l_2 m_2} = < a_{l_1 m_1} a_{l_2 m_2} >$ and $S$ is the normalization factor. In real calculations, it is difficult to obtain $C_{(a)}^{-1}$ if inhomogeneous noise and cut sky are considered, since $C_{(a)}$ is a huge matrix when $l$ is large, it is impossible to find its inverse directly. But once $C_{(a)}^{-1}a$ can be quickly calculated, a feasible algorithm based on this estimator can be executed \cite{Smith:2006ud}. Nowadays, $C_{(a)}^{-1}a$ is approximated by $\frac{a_{lm}}{C_l}$ \cite{Komatsu:2003iq,Creminelli:2005hu,Yadav:2007ny}, which is easy to calculate. After this approximation the estimator becomes,
\begin{eqnarray}
\epsilon_{lin}(a) &=& \frac{1}{S} \sum_{(l,m)} [\frac {(< a_{l_1 m_1} a_{l_2 m_2} a_{l_3 m_3} >_{f_{NL}=1})}{C_{l_1} C_{l_2} C_{l_3}} a_{l_1 m_1} a_{l_2 m_2} a_{l_3 m_3}   \nonumber\\
&& -3 \frac {(< a_{l_1 m_1} a_{l_2 m_2} a_{l_3 m_3} >_{f_{NL}=1})}{C_{l_1} C_{l_2} C_{l_3}} C_{(a)l_1 m_1,l_2 m_2}  a_{l_3 m_3}].
\label{eaa}
\end{eqnarray}
Besides in the non-Gaussianity estimation, the same problem also arise in power spectrum estimation \cite{Oh:1998sr} and CMB lensing detection \cite{Smith:2007rg}.

Although the inhomogeneous noise and cut sky effect is already considered in the estimator (\ref{ea}, \ref{eaa}), it is incomplete, since the noise and sky effect enter into the estimator only through the spherical harmonical coefficient $a_{lm}$. The aim of the present paper is to give a complete treatment of the inhomogeneous noise. We calculate the probability distribution function (PDF) of CMB anisotropy in the inhomogeneous noise and cut sky case. From the PDF a new estimator is obtained. For slightly inhomogeneous noise it is possible to compute the estimator order by order. Each order is suppressed by two factors: $(\frac{\sigma^2}{\sigma^2_{i}}-1)$ and $\frac{C_l}{C^{tot}_{l}}$, so the series converges.

This paper is organized as follows. In section II, the PDF for the CMB anisotropy is considered, from which we obtaine the optimal estimator for local type of non-Gaussianity. In section III, we expand the new estimator, and get a series expansion method to deal with the inhomogeneous noise. Through a detailed analysis we show that in what condition our estimator reduces to the popularly employed one and how our method reflects the inhomogeneous effect. In the last section, we first generalize the estimator to other types of non-Gaussianity, then we also discuss the slightly inhomogeneous condition and the noise weight. Appendix A is devoted to the relations appearing in the radiative transfer process, while Appendix B is devoted to the normalization factor.
\section{The Bispectrum Estimator}
%\subsection{general case}
The observed CMB anisotropy is composed of several components: the primordial temperature fluctuation, the foreground emission, secondary effects and the instrument noise. In this paper we do not consider the foreground emission and secondary effects, since the foreground emission has already been subtracted \cite{Bennett:2003bz} and secondary effects such as point sources and CMB lensing are carefully studied in \cite{Serra:2008wc,Babich:2008uw,Cooray:2008xz}. Therefore each pixelied CMB data can be regarded as superposition of the CMB signal and the instrument noise,
\begin{eqnarray}
\delta T_{i}=\delta T^{s}_{i}+\delta T^{n}_{i}.
\end{eqnarray}
For the $i_{th}$ pixel, the total PDF is just a simple convolution of the PDF of the signal and the PDF of the noise \cite{Jeong:2007mx}:
\begin{eqnarray}
f(\delta T_{i})=\int d\delta T^{s}_{i} d\delta T^{n}_{i} f_{s}(\delta T^{s}_{i})f_{n}(\delta T^{n}_{i}) \delta_{D}(\delta T_{i}-\delta T^{s}_{i}-\delta T^{n}_{i}),
\label{f}
\end{eqnarray}
where $f_{s}$ is the PDF of the signal, $f_{n}$ are the PDF of the noise and $\delta_D$ is the Kronecker delta function. The instrument noise for each pixel is always assumed to follow the gaussian distribution with a variance $\sigma_{i}$,
\begin{equation}
f_n(\delta T^n_i)=\frac{1}{\sqrt{2\pi\sigma_i^2}}\exp[-\frac{(\delta T^{n}_i)^2}{2\sigma_i^2}],
\label{fn}
\end{equation}
where $\sigma_{i}$ varies from pixel to pixel. For a given primordial curvature perturbations $\Phi(\vec x)$, the PDF of the signal $f_{s}(\delta T^{s}_{i})$ can be written as:
\begin{eqnarray}
f_{s}(\delta T^{s}_{i})=\int d^{N}\Phi \delta_{D}(\delta T_{i}^{s}-\int r^{2} dr \sum_{l} \alpha_{l}(r)b_{l} \sum_{m} \Phi_{lm}(r)Y_{lm}(\vec n_{i})) f(\Phi),
\label{fs}
\end{eqnarray}
where $b_l$ is the window function which reflects the beam and pixel smearing, $f(\Phi)$ is the PDF of the curvature perturbation and $\alpha_{l}(r)$ is defined by EQ (\ref{al}). $\Phi_{lm}(r)$ is the spherical harmonic coefficient of the curvature perturbations at a given distance $r$ and $\vec n_i$ is the unit vector pointing to the $i_{th}$ pixel. For the local type of primordial non-Gaussianity (\ref{phi}), the PDF $f(\Phi)$ is given by:
\begin{equation}
f(\Phi)=\int d^{N}\Phi_L \delta^{N}_D(\Phi(\vec x)-\Phi_L(\vec x)-f_{NL}[\Phi^2_L(\vec x)-<\Phi^2_L(\vec x)>])\frac{\exp[-\frac{1}{2}\Phi_LC_{\Phi}^{-1}\Phi_L]}{\sqrt{(2\pi)^NdetC_{\Phi}}},
\label{fphi}
\end{equation}
where the functional $f(\Phi)$ is written in compact vector notation. In the spherical harmonical space, $C_{\Phi}$ is given by EQ(\ref{cphi}). Substituting EQ (\ref{fn}), (\ref{fs}), and (\ref{fphi}) into EQ (\ref{f}), we obtain the one point PDF for the $i_{th}$ pixel,
\begin{eqnarray}
f(\delta T_{i}) &=& \int d^{N}\Phi d^{N}\Phi_L d\delta T^{n}_{i} \delta_{D}(\delta T_{i}-\int r^{2} dr \sum_{l} \alpha_{l}(r)b_{l} \sum_{m} \Phi_{lm}(r)Y_{lm}(\vec n_{i})-\delta T^{n}_{i}) \nonumber\\
& & \frac{1}{\sqrt{2\pi\sigma_i^2}}\exp[-\frac{(\delta T^n_i)^2}{2\sigma_i^2}]\delta^{N}_D(\Phi(\vec x)-\Phi_L(\vec x)-f_{NL}[\Phi^2_L(\vec x)-<\Phi^2_L(\vec x)>])\frac{\exp[-\frac{1}{2}\Phi_LC_{\Phi}^{-1}\Phi_L]}{\sqrt{(2\pi)^NdetC_{\Phi}}}.
\end{eqnarray}
The above formula can be easily generalized to the united PDF for more pixels,
\begin{eqnarray}
f(\delta T_{1},\cdots,\delta T_{m}) &=& \int d^{N}\Phi d^{N}\Phi_L \prod^m_{i=1} \{ d\delta T^{n}_{i} \delta_{D}(\delta T_{i}-\int r^{2} dr \sum_{l} \alpha_{l}(r)b_{l} \sum_{m} \Phi_{lm}(r)Y_{lm}(n_{i})-\delta T^{n}_{i}) \nonumber\\
& & \frac{1}{\sqrt{2\pi\sigma_i^2}}\exp[-\frac{(\delta T^n_i)^2}{2\sigma_i^2}] \} \delta^{N}_D(\Phi(\vec x)-\Phi_L(\vec x)-f_{NL}[\Phi^2_L(\vec x)-<\Phi^2_L(\vec x)>])\frac{\exp[-\frac{1}{2}\Phi_LC_{\Phi}^{-1}\Phi_L]}{\sqrt{(2\pi)^NdetC_{\Phi}}}.
\label{f1m}
\end{eqnarray}
Today's CMB constraint indicates that the primordial non-Gaussianity is weak, so we can calculate the PDF (\ref{fphi}) to $\mathcal{O}(f_{NL}\mu)$ ($\mu = <\Phi^2(x)>$) and safely ignore higher order terms. This is already done in \cite{Babich:2005en}. Here we just present the final result,
\begin{eqnarray}
f(\Phi)& = & \exp [-\frac{1}{2} \sum_{l,m} \int r_1^2 dr_1 r_2^2 dr_2 
   \Phi^*_{l m}(r_1) D^{-1}_{l}(r_1, r_2) \Phi_{l m}(r_2) \nonumber \\
& + & f_{NL} \sum_{(l,m)} \int r_1^2 dr_1 r_2^2 dr_2 \mathcal{G}^{l_1 l_2 l_3}_{m_1 m_2 m_3}
   \Phi_{l_1 m_1}(r_1) D^{-1}_{l_1}(r_1,r_2) \Phi_{l_2 m_2}(r_2) \Phi_{l_3 m_3}(r_2) + \mathcal{O}(f^2_{NL}\mu^2)], 
\end{eqnarray}
where $D^{-1}_{l}(r_1,r_2)$ is defined by EQ (\ref{dinverse}), and $\mathcal{G}^{l_1 l_2 l_3}_{m_1 m_2 m_3}$ is Gaunt Integral defined by EQ (\ref{gaunt}). Then substitute this result into EQ (\ref{f1m}) and do the $\delta T$ integral, the PDF (\ref{f1m}) is simplified to a Gaussian functional integral,
\begin{eqnarray}
f(\delta T_{1},\cdots,\delta T_{m}) &=& \exp[f_{NL} \sum_{(l,m)} \int r_1^2 dr_1 r_2^2 dr_2 \mathcal{G}^{l_1 l_2 l_3}_{m_1 m_2 m_3} \frac{\partial}{\partial J^{*}_{l_1 m_1}(r_1)} D^{-1}_{l_1}(r_1,r_2) \frac{\partial}{\partial J^{*}_{l_2 m_2}(r_2)} \frac{\partial}{\partial J^{*}_{l_3 m_3}(r_2)} ]\nonumber\\ 
&& \{ \int d^{N}\Phi_{lm} d^{N}\Phi^*_{lm} (\prod_i \frac{1}{\sqrt{2\pi\sigma_i^2}}) \exp[-\sum_i \frac{\delta T^2_{i}}{2\sigma_i^2}] \nonumber\\
&& \exp[-\Phi^* C \Phi+ J^* \Phi+ J \Phi^*+A^* \Phi+ A \Phi^*] \},
\end{eqnarray}
with
\begin{eqnarray}
C_{lmr,(lmr)^{'}} = \frac{r^2r^{'2}}{2}(D^{-1}_{l}(r,r^{'})\delta_{ll^{'}}\delta_{mm^{'}}+\alpha_l(r) b_l \alpha_{l^{'}}(r^{'}) b_{l^{'}} N^{-1}_{lm,(lm)^{'}})
\end{eqnarray}
\begin{eqnarray}
N^{-1}_{lm,(lm)^{'}} = \sum_{i=1}^{N_{pix}} \frac{Y^*_{lm}(\vec n_{i})Y_{l^{'}m^{'}}(\vec n_{i})}{\sigma^2_{i}} M(\vec n_i)
\end{eqnarray}
\begin{eqnarray}
A_{lm}(r) = \frac{r^2 \alpha_l(r) b_l}{2}\sum_{i=1}^{N_{pix}} Y^*_{lm}(\vec n_{i}) \frac{\delta T_{i}}{\sigma_i^2} M(\vec n_i),
\end{eqnarray}
where $M(\vec n_i)$ is the mask function, which take value unity when $i=1,\ldots ,m$ and zero in other case. Here and in the following we use the following abbreviations, $\Phi^* C \Phi\equiv\int dr dr^{'} \sum_{(lml^{'}m^{'})} \Phi_{l m}^*(r) C_{lmr,(lmr)^{'}} \Phi_{l^{'} m^{'}}(r^{'})$, $J^* \Phi\equiv\int dr \sum_{lm} J^*_{lm}(r) \Phi_{l m}(r)$ and $A^* \Phi\equiv\int dr \sum_{lm} A^*_{lm}(r) \Phi_{l m}(r)$. After performing the integral over $\Phi$ and dropping the irrelevant constant factors, we obtain
\begin{eqnarray}
\lg f = A^* C^{-1} A+f_{NL} \sum_{(l,m)} \int r_1^2 dr_1 r_2^2 dr_2 \mathcal{G}^{l_1 l_2 l_3}_{m_1 m_2 m_3} (C^{-1}A)_{l_1 m_1}(r_1) D^{-1}_{l_1}(r_1,r_2) (C^{-1}A)_{l_2 m_2}(r_2) (C^{-1}A)_{l_3 m_3}(r_2).
\end{eqnarray}
It is obvious that performing the Gaussian integral is equivalent to simply substituting $\Phi_{lm}$ by $C^{-1}A$. Acoording to \cite{Babich:2005en}, the optimal estimator of $f_{NL}$ can be taken as:
\begin{eqnarray}
\epsilon (T) = \frac{1}{S} \sum_{(l,m)} \int r_1^2 dr_1 r_2^2 dr_2 \mathcal{G}^{l_1 l_2 l_3}_{m_1 m_2 m_3} (C^{-1}A)_{l_1 m_1}(r_1) D^{-1}_{l_1}(r_1,r_2) (C^{-1}A)_{l_2 m_2}(r_2) (C^{-1}A)_{l_3 m_3}(r_2),
\label{et}
\end{eqnarray}
where $S$ is the normalization factor.

EQ (\ref{et}) is our first main result. In deriving formula (\ref{et}) we do not require the condition of full sky and homogeneous noise, so EQ (\ref{et}) can be applied to cut sky and inhomogeneous noise. It should be noted that even if one only considers the cubic term (we do not consider linear term for the moment), EQ (\ref{et}) is also different from EQ (\ref{ea}).  $A$ in EQ (\ref{et}) corresponds to the ordinary spherical harmonical coefficient $a_{lm}$ in EQ (\ref{ea}), but it is noise and sky weighted. $C$ in EQ (\ref{et}) corresponds to $C_{(a)}$ in EQ (\ref{ea}), but $C$ is a function of $r$ while $C_{(a)}$ is not. This is because only the PDF of curvature perturbation is approximately considered to $f_{NL}\mu$ order, the cut sky and inhomogeneous noise effects for each pixel are rigorously accounted, while in writing down EQ (\ref{ea}) the cut sky and noise effects are only partially accounted through the ordinary spherical harmonical coefficient $a_{lm}$.
\section{The Inhomogeneous Noise}
Although EQ (\ref{et}) is a general estimator in the case of inhomogeneous noise and cut sky, it is difficult to calculate the inverse $C^{-1}A$, which becomes the main problem in the real computation \cite{Smith:2006ud}. Luckily there exists a series expansion method to deal with the inhomogeneous noise when the noise is slightly inhomogeneous. In order to establish the conventions, we first consider the case of full sky and homogenious noise and then turn into the case of cut sky and inhomogeneous noise. It must be stressed that the method presented in the following cannot be used to deal with the cut sky effect even there is no noise. In the subsequent section the window function $b_l$ has been absorbed into the definition of $\alpha_l,\beta_l,C_l$, therefore when $\alpha_l,\beta_l,C_l$ appear, they mean $\alpha_l b_l,\beta_l b_l,C_l b_l^2$.
\subsection{Full sky and homogeneous noise}
In the case of full sky and homogeneous noise, the noise level is the same for each pixel (denoted by $\sigma$), so the estimator (\ref{et}) will be greatly simplified. First note that in this case $N^{-1},C,A$ reduce to
\begin{eqnarray}
N^{-1}_{lm,(lm)^{'}} = \frac{1}{\sigma^2\frac{4\pi}{N_{pix}}}\delta_{ll^{'}}\delta_{mm^{'}},
\end{eqnarray}
\begin{eqnarray}
C_{lmr,(lmr)^{'}}=\frac{r^2r^{'2}}{2}(D^{-1}_{l}(r,r^{'})+\frac{\alpha_l(r)\alpha_l(r^{'})}{\sigma^2\frac{4\pi}{N_{pix}}})\delta_{ll^{'}}\delta_{mm^{'}},
\label{A0}
\end{eqnarray}
\begin{eqnarray}
A_{lm}(r)=\frac{r^2 \alpha_l(r)}{2\sigma^2\frac{4\pi}{N_{pix}}}a_{lm},
\end{eqnarray}
where $a_{lm} = \frac{4\pi}{N_{pix}} \sum^{N_{pix}}_{i=1} Y^{*}_{lm}(\vec n_i) \delta T_i$ is the ordinary spherical harmonical coefficient. It is easy to find the inverse $C^{-1}$
\begin{eqnarray}
(C^{-1})_{lmr}=2(D_{l}(r,r^{'})-\frac{\beta_l(r)\beta_l(r^{'})}{C^{tot}_l})\delta_{ll^{'}}\delta_{mm^{'}},
\label{A0inverse}
\end{eqnarray}
where $C^{tot}_l=C_l+\sigma^2\frac{4\pi}{N_{pix}}$ is the total angular power spectrum. Then $C^{-1}A$ becomes
\begin{eqnarray}
(C^{-1}A)_{lmr}=\frac{\beta_l(r)}{C^{tot}_l}a_{lm},
\end{eqnarray}
which gives the optimal estimator,
\begin{eqnarray}
\epsilon (T) = \frac{1}{S} \sum_{(l,m)} \int r^2 dr \mathcal{G}^{l_1 l_2 l_3}_{m_1 m_2 m_3} \frac{\alpha_{l_1}(r)}{C^{tot}_{l_1}}a_{l_1 m_1}  \frac{\beta_{l_2}(r)}{C^{tot}_{l_2}}a_{l_2 m_2} \frac{\beta_{l_3}(r)}{C^{tot}_{l_3}}a_{l_3 m_3}.
\label{eu}
\end{eqnarray}
This is the estimator used in \cite{Komatsu:2003iq}. So for full sky and homogeneous noise, our estimator (\ref{et}) reduces to the ordinary one, as is expected. 
\subsection{Cut sky and inhomogeneous noise}
In the case of cut sky and inhomogeneous noise, the weight (including the sky mask and the noise level) is different for each pixel, so it is impossible to calculate the inverse of $C$ directly. However in the case of slightly inhomogeneous noise, there exists a simple way to calculate $C^{-1}A$. In this case, we split the matrix $N^{-1}$ into two parts:
\begin{eqnarray}
N^{-1}_{lm,(lm)^{'}} &=& \sum^{N_{pix}}_{i=1} \frac{Y^*_{lm}(n_{i})Y_{l^{'}m^{'}}(n_{i})}{\sigma^2}(1+\frac{\sigma^2}{\sigma^2_{i}}M(n_i)-1)\nonumber\\
&=&\frac{1}{\sigma^2\frac{4\pi}{N_{pix}}}\delta_{ll^{'}}\delta_{mm^{'}}+N^{-1S}_{lm,(lm)^{'}}
\label{ff}
\end{eqnarray}
with
\begin{eqnarray}
\frac{1}{\sigma^2}=\frac{1}{N} \sum^{N}_{i=1}\frac{1}{\sigma^2_{i}}
\label{sigma}
\end{eqnarray}
\begin{eqnarray}
N^{-1S}_{lm,(lm)^{'}}=\frac{1}{\sigma^2} \sum^{N_{pix}}_{i=1} Y^*_{lm}(\vec n_{i})Y_{l^{'}m^{'}}(\vec n_{i})(\frac{\sigma^2}{\sigma^2_{i}}M(\vec n_i)-1),
\label{ns}
\end{eqnarray}
where $N$ is the total number of the unmasked pixels. The first term in the last line of EQ (\ref{ff}) represents an average effect, and the second term is a small perturbation about the average. Correspondingly, $A$ can also be split into two parts:
\begin{eqnarray}
C_{lmr,(lmr)^{'}} &=& \frac{r^2r^{'2}}{2}(D^{-1}_{l}(r,r^{'})+\frac{\alpha_l(r)\alpha_{l^{'}}(r^{'})}{\sigma^2\frac{4\pi}{N_{pix}}})\delta_{ll^{'}}\delta_{mm^{'}}+\frac{r^2\alpha_l(r)r^{'2}\alpha_{l^{'}}(r^{'})}{2}N^{-1S}_{lm,(lm)^{'}} \nonumber\\
&\equiv& (C^{(0)})_{lmr,(lmr)^{'}} +(B)_{lmr,(lmr)^{'}}.
\label{cseries}
\end{eqnarray}
The first term is just EQ (\ref{A0}), its inverse is given by EQ (\ref{A0inverse}). Compared with the first term, the last term is small, so we can calculate the inverse $C^{-1}$ order by order,
\begin{eqnarray}
C^{-1} = C^{(0)-1} +C^{(1)}+C^{(2)}+\cdots+C^{(n)}+\cdots,
\label{ccc}
\end{eqnarray}
where $C^{n}$ is given by $C^{(n)} = (-1)^n (C^{(0)-1}B)^n C^{(0)-1}$. From EQ (\ref{cseries}), $C^{(0)-1}B$ is represented as
\begin{eqnarray}
(C^{(0)-1}B)_{lmr,(lmr)^{'}} = \frac{\beta_{l}(r)}{C^{tot}_{l}} \sigma^2\frac{4\pi}{N_{pix}}  N^{-1S}_{lm,(lm)^{'}} r^{'2} \alpha_{l^{'}}(r),
\label{bainverse}
\end{eqnarray}
and $C^{(0)-1}A$ is given by
\begin{eqnarray}
(C^{(0)-1}A)_{lmr} = \frac{\beta_l(r)}{C^{tot}_l} (\sigma^2\frac{4\pi}{N_{pix}} b_{lm}),
\label{caaa}
\end{eqnarray}
with
\begin{eqnarray}
b_{lm}&=&\sum^{N_{pix}}_{i=1} Y^*_{lm}(\vec n_{i}) \frac{\delta T_{i}}{\sigma_i^2}M(\vec n_i).
\label{blm}
\end{eqnarray}
Thus, combining with EQ (\ref{ccc}), EQ (\ref{bainverse}) and EQ (\ref{caaa}),  $(C^{-1}A)_{lmr}$ can be calculated by the series:
\begin{eqnarray}
(C^{-1}A)_{lmr} &=& (C^{(0)-1}A)_{lmr} + \cdots+ (-1)^n ((C^{(0)-1}B)^n C^{(0)-1}A)_{lmr} + \cdots  \nonumber\\
&=& \frac{\beta_l(r)}{C^{tot}_l} (\sigma^2\frac{4\pi}{N_{pix}}b_{lm}) - \frac{\beta_l(r)}{C^{tot}_l} \sum_{l_1 m_1} (\sigma^2 \frac{4\pi}{N_{pix}} N^{-1S}_{lm,l_1 m_1}) \frac{C_{l_1}}{C^{tot}_{l_1}} (\sigma^2\frac{4\pi}{N_{pix}}b_{l_1 m_1}) \nonumber\\
&+& \frac{\beta_l(r)}{C^{tot}_l} \sum_{l_1 m_1 l_2 m_2} (\sigma^2 \frac{4\pi}{N_{pix}} N^{-1S}_{lm,l_1 m_1}) \frac{C_{l_1}}{C^{tot}_{l_1}} (\sigma^2 \frac{4\pi}{N_{pix}} N^{-1S}_{l_1 m_1,l_2 m_2}) \frac{C_{l_2}}{C^{tot}_{l_2}} (\sigma^2\frac{4\pi}{N_{pix}}b_{l_2 m_2}) + \cdots  \nonumber\\
&\equiv& \frac{\beta_l(r)}{C^{tot}_l} \sigma^2 \frac{4\pi}{N_{pix}} (b_{lm}+\sum_{l^{'} m^{'}}F_{lm,l^{'} m^{'}} b_{l^{'}m^{'}}) \nonumber\\
&\equiv& \frac{\beta_l(r)}{C^{tot}_l} T_{lm}.
\label{Aic}
\end{eqnarray}
In last second line we formally define the matrix $F$ which will be used in Appendix B and in the last line we define a quantity $T_{lm}$ which is analogous to the spherical harmonical coefficient $a_{lm}$. With $T_{lm}$ the estimator  (\ref{et}) can be written in a compact form,
\begin{eqnarray}
\epsilon (T) = \frac{1}{S} \sum_{(l,m)} \int r^2 dr \mathcal{G}^{l_1 l_2 l_3}_{m_1 m_2 m_3} \frac{\alpha_{l_1}(r)}{C^{tot}_{l_1}}T_{l_1 m_1}  \frac{\beta_{l_2}(r)}{C^{tot}_{l_2}}T_{l_2 m_2} \frac{\beta_{l_3}(r)}{C^{tot}_{l_3}}T_{l_3 m_3}.
\label{estimator}
\end{eqnarray}
This is our second main result. Compared with the original one (\ref{et}), now the estimator has the same form as the ordinary one (\ref{eu}), the only change is substituting $a_{lm}$ by $T_{lm}$, which is calculated by the seires (\ref{Aic}). Each order in the series can be computed as easily as $a_{lm}$. In the following two subsections it will be shown how $T_{lm}$ takes the inhomogeneous noise effect into account.
\subsection{Full sky and inhomogeneous noise}
In this subsection we consider the case of slightly inhomogeneous noise but full sky. Under this condition, $M(n_i)$ in $N^{-1}_{lm,(lm)^{'}}$ (\ref{ns}), $b_{lm}$ (\ref{blm}) all equal unity and $\sigma$ (\ref{sigma}) is determined by $\frac{1}{\sigma^2}=\frac{1}{N_{pix}} \sum^{N_{pix}}_{i=1}\frac{1}{\sigma^2_{i}}$. From EQ (\ref{Aic}), we see that there are two factors to suppress the series. The first one is the factor $\frac{\sigma^2}{\sigma^2_{i}}-1$ appearing in $N^{-1S}_{lm,(lm)^{'}}$, and the other is the factor $\frac{C_{l}}{C^{tot}_{l}}$. In the large scale region the latter factor is close to unity, however, with $l$ increasing, where noise begins to dominate, it becomes smaller and smaller. The first factor does not vary with $l$, it is determined by experiment and small as long as the slightly inhomogeneous condition is satisfied. Therefore, the effect of the inhomogeneous noise can be accounted order by order through (\ref{Aic}). 

In order to show what $T_{lm}$ is compared with $a_{lm}$, we keep $C^{-1}A$ (\ref{Aic}) to first order
\begin{eqnarray}
(C^{-1}A)_{lmr} & = & \frac{\beta_l(r)}{C^{tot}_l} T_{lm} \nonumber\\
& = & \frac{\beta_l(r)}{C^{tot}_l} \sigma^2\frac{4\pi}{N_{pix}}(b_{lm}-\sum_{l^{'}m^{'}} N^{-1S}_{lm,(lm)^{'}} (\frac{C_{l^{'}}}{C^{tot}_{l^{'}}}) (\sigma^2\frac{4\pi}{N_{pix}}b_{l^{'}m^{'}})),
\label{ca1}
\end{eqnarray}
and split $b_{lm}$ into two terms
\begin{eqnarray}
b_{lm} = \frac{1}{\sigma^2\frac{4\pi}{N_{pix}}} (a_{lm}+ d_{lm}),
\end{eqnarray}
with $a_{lm} = \frac{4\pi}{N_{pix}} \sum^{N_{pix}}_{i=1} Y^*_{lm}(\vec n_{i}) \delta T_{i}$ and $d_{lm} = \frac{4\pi}{N_{pix}} \sum^{N_{pix}}_{i=1} Y^*_{lm}(\vec n_{i}) \delta T_{i} (\frac{\sigma^2}{\sigma^2_{i}}-1)$. Compared with $a_{lm}$, $d_{lm}$ is a small quantity. Pluging $a_{lm}$ and $d_{lm}$ into EQ (\ref{ca1}) and ignoring $d_{lm}$ in the last term in EQ (\ref{ca1}), we obtain 
\begin{eqnarray}
T_{lm} &=&a_{lm}+\frac{4\pi}{N_{pix}} \sum^{N_{pix}}_{i=1} Y^*_{lm}(\vec n_{i}) \delta T_{i} (\frac{\sigma^2}{\sigma^2_{i}}-1)+\frac{4\pi}{N_{pix}} \sum^{N_{pix}}_{i=1} Y^*_{lm}(\vec n_{i}) (\frac{\sigma^2}{\sigma^2_{i}}-1) \sum_{l^{'} m^{'}} Y_{l^{'} m^{'}}(\vec n_{i}) \frac{C_{l^{'}}}{C^{tot}_{l^{'}}} a_{l^{'}m^{'}} \nonumber\\
&=& a_{lm}+\frac{4\pi}{N_{pix}} \sum^{N_{pix}}_{i=1} Y^*_{lm}(\vec n_{i}) (\frac{\sigma^2}{\sigma^2_{i}}-1) \sum_{l^{'} m^{'}} Y_{l^{'} m^{'}}(\vec n_{i}) \frac{\sigma^2\frac{4\pi}{N_{pix}}}{C^{tot}_{l^{'}}} a_{l^{'}m^{'}}.
\label{tlm}
\end{eqnarray}
It is easy to see that the dominant part of $T_{lm}$ is the ordinary spherical harmonical coefficient $a_{lm}$. Besides this term, there is another small term which is suppressed by factors $\frac{\sigma^2\frac{4\pi}{N_{pix}}}{C^{tot}_{l}}$ and $\frac{\sigma^2}{\sigma^2_{i}}-1$. This term reflects the inhomogeneous noise. It should be pointed out that the factor $\frac{\sigma^2\frac{4\pi}{N_{pix}}}{C^{tot}_{l}}$ appears in the difference $T_{lm}-a_{lm}$, while the factor $\frac{C_l}{C^{tot}_{l}}$ appears in the series (\ref{Aic}) employed to calculate $T_{lm}$. So in small $l$ region, where the CMB signal dominates, $T_{lm}$ is close to $a_{lm}$, but the series (\ref{Aic}) converges slowly. While in the large $l$ region, $T_{lm}$ is different from $a_{lm}$, but the series converges rapidly.
\subsection{Cut sky and homogeneous noise}
In the previous section the case of full sky but inhomogeneous noise has already been discussed. Here we mainly discuss the case of cut sky but homogeneous noise. In this case, the second term of the series (\ref{Aic}) becomes
\begin{eqnarray}
&&\sum_{l_1 m_1} (\sigma^2 \frac{4\pi}{N_{pix}} N^{-1S}_{lm,l_1 m_1}) \frac{C_{l_1}}{C^{tot}_{l_1}} (\sigma^2 \frac{4\pi}{N_{pix}} b_{l_1 m_1}) \nonumber\\
&=& \frac{4\pi}{N_{pix}} \sum^{N_{pix}}_{i=1} Y^*_{lm}(\vec n_{i}) (M(\vec n_i)-1) \sum_{l_1 m_1} Y_{l_1 m_1}(\vec n_{i}) \frac{C_{l_1}}{C^{tot}_{l_1}} \frac{4\pi}{N_{pix}} \sum^{N_{pix}}_{j=1} Y^*_{l_1 m_1}(\vec n_{j}) \delta T_{j} M(\vec n_j),
\label{Ac}
\end{eqnarray}
where $(\sigma^2 \frac{4\pi}{N_{pix}} b_{l_1 m_1})=\frac{4\pi}{N_{pix}} \sum^{N_{pix}}_{j=1} Y^*_{l_1 m_1}(\vec n_{j}) \delta T_{j} M(\vec n_j)$ represents the temperature fluctuation of the unmasked region. In the small $l$ region, where the signal dominates, $\frac{C_{l}}{C^{tot}_{l}} \approx 1$, EQ (\ref{Ac}) becomes
\begin{eqnarray}
&&\frac{4\pi}{N_{pix}} \sum^{N_{pix}}_{i=1} Y^*_{lm}(\vec n_{i}) (M(\vec n_i)-1) \frac{4\pi}{N_{pix}} \sum^{N_{pix}}_{j=1} \delta T_{j} M(\vec n_j) \delta_{Dij} (\frac{4\pi}{N_{pix}})^{-1} \nonumber\\
&=& \frac{4\pi}{N_{pix}} \sum^{N_{pix}}_{ij=1} Y^*_{lm}(\vec n_{i}) \delta T_{i} M(\vec n_j) (M(\vec n_i)-1) \delta_{Dij} \nonumber\\
&=& 0 
\end{eqnarray}
In the large $l$ region where the noise dominates, $\frac{C_{l}}{C^{tot}_{l}}$ is small, so EQ (\ref{Ac}) is greatly suppressed. Therefore, in this case the second term in (\ref{Aic}) is also small compared with the first term.

However, we must stress again that the series (\ref{Aic}) can only be employed to treat the effect of noise, but failed to account for the effect of cut sky. This is because in the zero noise limit, $\frac{C_{l}}{C^{tot}_{l}}$ strictly equals unity for all $l$, all terms in the series (\ref{Aic}) are zero except the first one.
\subsection{The normalization factor}
The last problem is the normalization factor $S$. $S$ in EQ (\ref{estimator}) can be calculated formally order by order. This calculation is presented in Appendix B, but the form presented there is not suitable for numerical computation. So we need a fast algorithm to compute the normalization factor $S$. However the advantage of estimator (\ref{estimator}) is that the normalization factor needs to be computed only one time, while the part left which needs to be calculated repeatedly (to obtain the confident region of $f_{NL}$) is easy to compute.
\section{Discussion and conclusion}
(1) Until now we only consider the estimator for local type of non-Gaussianity. It is straightforward to generalize our results to other type of non-Gaussianity. Generally speaking, the PDF of non-Gaussianity curvature perturbation can be expressed by the Edgeworth expansion. The Edgeworth expansion of a multivariate PDF can be found in \cite{Taylor:2000hq} and \cite{Bernardeau:1994aq}. EQ (7) of \cite{Taylor:2000hq} indicates that the PDF of the curvature perturbation can be represented by,
\begin{eqnarray}
f(\Phi)=(1-X)f_0(\Phi),
\end{eqnarray}
where $f_0(\Phi)$ is the Gaussian part of the PDF and $X$ is a small perturbation which is function of $\Phi_{lm}$ (the exact form of $X$ for general non-Gaussianity can be found in \cite{Taylor:2000hq}). Following the same steps of section II, where the key point is to substitute $\Phi_{lm}$ in $X$ by $C^{-1}A$, we obtain the estimator for general non-Gaussianity,
\begin{eqnarray}
\epsilon(T)=\frac{1}{S} X(C^{-1}A),
\end{eqnarray}
and the linear term is already naturally included in the estimator.

(2) Now let us discuss the slightly inhomogeneous condition. Here we simply take the WMAP 3-year $V2$ band as an example. The noise level for each pixel is determined by the observation times $\sigma^2_{i} = \frac{\sigma^2_0}{N_{obs}(i)}$, $\sigma^2  = \frac{\sigma^2_0}{N_{eff}}$, where $\sigma^2_0$ is the rms noise per observation. From EQ (\ref{sigma}) we see that the effective observation times corresponding to the average noise level is $N_{eff} =\frac{1}{N} \sum_{i=1}^{N} N_{obs}(i)$. Since there are about $4 \%$ pixels with $N_{obs} > 1200 $, $1.8 \%$ pixels with $N_{obs} < 400 $ and for most of pixels the observation times is distributed in the region $[500, 1000]$, the factor $(\frac{\sigma^2}{\sigma^2_{i}}-1) = (\frac{N_{obs}(i)}{N_{eff}}-1)$ will be small for most pixels. Those pixels with too large or too small observation times can be masked out in the non-Gaussianity estimation.

As a matter of fact, there is no requirment that $\sigma^2$ must take the form of (\ref{sigma}), It can be any values. However in order to make the series calculation efficient, EQ (\ref{sigma}) is a good choice. There are two factors to affect the choice of $\sigma^2$, $(\frac{\sigma^2}{\sigma^2_{i}}-1)$ and $\frac{C_l}{C^{tot}_{l}}$. If $\sigma^2$ is too large or too small, the series (\ref{Aic}) will converge slowly.

(3) In preparing our paper, WMAP Group release their five-year data \cite{Komatsu:2008hk}. In order to deal with the inhomogeneous noise, they adopt the average noise and ``combination signal-plus-noise'' weight explicitly in EQ (A6), EQ (A18), EQ (A19) and EQ (A27) of \cite{Komatsu:2008hk}. Using the symbol conventions in our paper, they are $\sigma_1^2=\frac{1}{N} \sum_{i=1}^{N_{pix}} \sigma^2_{i} M(\vec n_i)$, $(\frac{1}{\sigma_{cmb}^2+\sigma_2^2})^3=\frac{1}{N} \sum^{N_{pix}}_{i=1} (\frac{M(\vec n_i)}{\sigma_{cmb}^2+\sigma_2^2})^3$, $W(\vec n_i)=\frac{M(\vec n_i)}{\sigma^2_{cmb}+\sigma^2_{i}}$ and $a_{lm} = \frac{4\pi}{N_{pix}} \sum_{i=1}^{N_{pix}} Y_{lm}(\vec n_i) \delta T_i M(\vec n_i) \frac{\sigma^2_{cmb}+\sigma_2^2}{\sigma^2_{cmb}+\sigma^2_{i}}$ (we already reexpress their equations for the present purpose), where $\sigma^2_{cmb} \equiv \frac{1}{4\pi} \sum_l (2l+1) C_l$ (remember that $b_l$ is already absorbed in the definition of $C_l$). In \cite{Komatsu:2008hk} $\sigma_1$ is used to calculate the total angular power spectrum $C_l^{tot} = C_l + \frac{4\pi}{N_{pix}} \sigma_1^2$, while $\sigma_2$ is the average noise appearing in the ``combination signal-plus-noise'' weight, they are not the same. Besides this, the noise weight is also different from the way used in our paper. We do not know why they use such a ``combination signal-plus-noise'' weight, but our discussion indicates that the inhomogeneous noise effect will be accounted better by the method presented in previous section, where the inhomogeneous noise is treated seriously.

In conclusion we have studied the inhomogeneous noise effect in the CMB non-Gaussianity estimation. First, we calculate the CMB anisotropy PDF for local type of non-Gaussianity which gives an optimal bispectrum estimator (\ref{et}). The estimator obtained in this paper is different from the ordinary one, since the cut sky and inhomogeneous noise factors are completely accounted in the PDF calculation, which provides a framework to study the effects of the cut sky and inhomogeneous noise in the CMB non-Gaussianity estimation. Then, we propose a series expansion method to calculate the new estimator. The final result is just to replace $a_{lm}$ which appears in the popular estimator (\ref{eu}) by $T_{lm}$. $T_{lm}$ can be calculated by the series (\ref{Aic}) as easily as $a_{lm}$. To the zeroth order, $T_{lm}$ reduces to $a_{lm}$ and the estimator (\ref{estimator}) becomes the usual one. Last, we discuss how to generalize the estimator (\ref{et}) to other type of non-Gaussianity, the slightly inhomogeneous condition and the noise weight.

\begin{acknowledgments}
We would like to thank Dr Sun weimin for improving the manuscript. This work was supported in part by the National Natural Science Foundation of China (under Grant No 10473023).
\end{acknowledgments}

\appendix
\section{Radiative transfer}
This appendix summarizes the relations which appears in the radiative transfer process. The contents are mainly based on section IV of \cite{Babich:2005en}.

In the spherical harmonic space, a given curvature perturbation $\Phi (\vec x)$ is represented by $\Phi_{lm} (r)$ . The covariance matrix of $\Phi_{lm} (r)$ can be calculated by
\begin{eqnarray}
\langle \Phi^{*}_{l_1 m_1}(r_1) \Phi_{l_2 m_2}(r_2) \rangle &=& \delta_{l_1 l_2} \delta_{m_1 m_2} D_{l_1}(r_1,r_2) 
\label{cphi} \\
	&=& \delta_{l_1 l_2} \delta_{m_1 m_2} \frac{2}{\pi} \int k^2 dk P(k) j_{l_1}(kr_1) j_{l_1}(kr_2).
\end{eqnarray}
where $P(k)$ is the power spectrum. Symbolically one defines the inverse of $D_{l}(r_1,r_2)$ as
\begin{equation}
  \int r^2 dr D_{l}(r_1,r) D^{-1}_{l}(r,r_2) = \frac{\delta(r_1 - r_2)}{r^2_1}.
\label{dinverse}
\end{equation}
Introduce the symbols:
\begin{equation}
  \alpha_l(r) = \frac{2}{\pi} \int k^2 dk j_l(kr) \Delta_l(k),
\label{al}
\end{equation}
\begin{equation}
  \beta_l(r) = \frac{2}{\pi} \int k^2 dk P(k) j_l(kr) \Delta_l(k),
\label{be}
\end{equation}
where $\Delta_l(k)$ is the radiation transfer fcuntion \cite{Ma:1995ey}. Then we obtain the useful formulas:
\begin{equation}
   \int r^2 dr \alpha_l(r) \beta_l(r) = C_l.
\label{cab}
\end{equation}
and
\begin{eqnarray}
  \alpha_l(r_1) = \int r^2_2 dr_2 D^{-1}_l(r_1, r_2) \beta_l(r_2), \\
\label{adb}
  \beta_l(r_1) = \int r^2_2 dr_2 D_l(r_1, r_2) \alpha_l(r_2).
\label{bda}
\end{eqnarray}
Finally we define the reduced bispectrum for local type of non-Gaussianity,
\begin{equation}
b_{l_1 l_2 l_3} = 2 \int r^2 dr [\alpha_{l_1}(r) \beta_{l_2}(r) \beta_{l_3}(r) + 
      \beta_{l_1}(r) \alpha_{l_2}(r) \beta_{l_3}(r) + \beta_{l_1}(r) \beta_{l_2}(r) \alpha_{l_3}(r)].
\label{reduceb} 
\end{equation}
When the window function is considered, EQ (\ref{cab}), (\ref{adb}), (\ref{bda}), (\ref{reduceb}) do not change except the replacement $\alpha_l(r) \rightarrow \alpha_l(r) b_l$, $\beta_l(r) \rightarrow \beta_l(r) b_l$, $C_l \rightarrow C_l b_l^2$, $b_{l_1 l_2 l_3} \rightarrow b_{l_1 l_2 l_3} b_{l_1} b_{l_2} b_{l_3}$.
\section{The normalization factor}
In this appendix we give the normalization factor formally. First, the 3-point correlator of $\delta T_{i}$ can be calculated by
\begin{eqnarray}
<\delta T_{i}\delta T_{j}\delta T_{k}> &=& <(\delta T^s_{i}+\delta T^n_{i})(\delta T^s_{j}+\delta T^n_{j})(\delta T^s_{k}+\delta T^n_{k})> \nonumber\\
&=& <\delta T^s_{i}\delta T^s_{j}\delta T^s_{k}> \nonumber\\
&=& <a_{l_1,m_1} a_{l_2,m_2} a_{l_3,m_3}> b_{l_1} b_{l_2} b_{l_3} Y_{l_1 m_1}(\vec n_{i}) Y_{l_2 m_2}(\vec n_{j}) Y_{l_3 m_3}(\vec n_{k}) \nonumber\\
&=& \mathcal{G}^{l_1 l_2 l_3}_{m_1 m_2 m_3} b_{l_1 l_2 l_3} Y_{l_1 m_1}(n_{i}) Y_{l_2 m_2}(n_{j}) Y_{l_3 m_3}(n_{k}),
\label{3t}
\end{eqnarray}
where $b_{l_1 l_2 l_3}$ is the reduced bispectrum defined by EQ (\ref{reduceb}) (the window function is already absorbed in the definition), $\mathcal{G}^{l_1 l_2 l_3}_{m_1 m_2 m_3}$ is the Gaunt Integral,
\begin{eqnarray}
\mathcal{G}^{l_1 l_2 l_3}_{m_1 m_2 m_3} = \int d^2 \vec{n} Y_{l_1,m_1} (\vec{n}) Y_{l_2,m_2} (\vec{n}) Y_{l_3,m_3} (\vec{n}).
\label{gaunt}
\end{eqnarray}
When deriving EQ (\ref{3t}), we have used the follwing formula
\begin{equation}
 \langle a_{l_1 m_1} a_{l_2 m_2} a_{l_3 m_3} \rangle = \mathcal{G}_{m_1 m_2 m_3}^{l_1 l_2 l_3} b_{l_1 l_2 l_3}.
\end{equation}
Then the 3-point correlator of $b_{lm}$ can be formally represented by
\begin{eqnarray}
<b_{l_1m_1}b_{l_2m_2}b_{l_3m_3}> &=& \sum^{N_{pix}}_{ijk=1} \frac{Y^*_{l_1m_1}(\vec n_{i})}{\sigma_i^2} \frac{Y^*_{l_2m_2}(\vec n_{j})}{\sigma_j^2} \frac{Y^*_{l_3m_3}(\vec n_{k})}{\sigma_k^2} M(\vec n_i)M(\vec n_j)M(\vec n_k) <\delta T_{i}\delta T_{j}\delta T_{k}> \nonumber\\
&=& \sum_{l^{'}m^{'}} \mathcal{G}^{l^{'}_1 l^{'}_2 l^{'}_3}_{m^{'}_1 m^{'}_2 m^{'}_3} b_{l^{'}_1 l^{'}_2 l^{'}_3} \sum^{N_{pix}}_{i=1} \frac{Y^*_{l_1 m_1}(\vec n_{i})Y_{l^{'}_1 m^{'}_1}(\vec n_{i})}{\sigma_i^2} M(\vec n_i) \sum^{N_{pix}}_{j=1} \frac{Y^*_{l_2 m_2}(\vec n_{j})Y_{l^{'}_2 m^{'}_2}(\vec n_{j})}{\sigma_j^2} M(\vec n_j) \nonumber\\
&& \sum^{N_{pix}}_{k=1} \frac{Y^*_{l_3 m_3}(\vec n_{k})Y_{l^{'}_3 m^{'}_3}(\vec n_{k})}{\sigma_k^2} M(\vec n_k) \nonumber\\
&\equiv& H^{l_1 l_2 l_3}_{m_1 m_2 m_3},
\label{H}
\end{eqnarray}
With the 3-point correlator of $b_{lm}$, the 3-point $T_{lm}$ can be formally written as
\begin{eqnarray}
<T_{l_1m_1}T_{l_2m_2}T_{l_3m_3}> &=& H^{l_1 l_2 l_3}_{m_1 m_2 m_3}   \nonumber\\
&+& [\sum_{l^{'} m^{'}}F_{l_1m_1,l^{'}_1 m^{'}_1} H^{l^{'}_1 l_2 l_3}_{m^{'}_1 m_2 m_3}+cycl.]  \nonumber\\
&+& [\sum_{l^{'} m^{'}} F_{l_1m_1,l^{'}_1 m^{'}_1} F_{l_2m_2,l^{'}_2 m^{'}_2} H^{l^{'}_1 l^{'}_2 l_3}_{m^{'}_1 m^{'}_2 m_3} +cycl.] \nonumber\\
&+& \sum_{l^{'} m^{'}} F_{l_1m_1,l^{'}_1 m^{'}_1} F_{l_2m_2,l^{'}_2 m^{'}_2} F_{l_3m_3,l^{'}_3 m^{'}_3} H^{l^{'}_1 l^{'}_2 l^{'}_3}_{m^{'}_1 m^{'}_2 m^{'}_3}.
\label{TTT} 
\end{eqnarray}
Where the matrix $F$ is defined in EQ (\ref{Aic}) and $H$ is defined in EQ (\ref{H}) . So the normalization factor $S$ can be formally represented by
\begin{eqnarray}
S = \sum_{(l,m)} \int r^2 dr \mathcal{G}^{l_1 l_2 l_3}_{m_1 m_2 m_3} \frac{\alpha_{l_1}(r)}{C^{tot}_{l_1}} \frac{\beta_{l_2}(r)}{C^{tot}_{l_2}} \frac{\beta_{l_3}(r)}{C^{tot}_{l_3}} <T_{l_1m_1}T_{l_2m_2}T_{l_3m_3}>.
\end{eqnarray} 
%\bibliography{noise}

\begin{thebibliography}{30}
\expandafter\ifx\csname natexlab\endcsname\relax\def\natexlab#1{#1}\fi
\expandafter\ifx\csname bibnamefont\endcsname\relax
  \def\bibnamefont#1{#1}\fi
\expandafter\ifx\csname bibfnamefont\endcsname\relax
  \def\bibfnamefont#1{#1}\fi
\expandafter\ifx\csname citenamefont\endcsname\relax
  \def\citenamefont#1{#1}\fi
\expandafter\ifx\csname url\endcsname\relax
  \def\url#1{\texttt{#1}}\fi
\expandafter\ifx\csname urlprefix\endcsname\relax\def\urlprefix{URL }\fi
\providecommand{\bibinfo}[2]{#2}
\providecommand{\eprint}[2][]{\url{#2}}

\bibitem[{\citenamefont{Spergel et~al.}(2003)}]{Spergel:2003cb}
\bibinfo{author}{\bibfnamefont{D.~N.} \bibnamefont{Spergel}}
  \bibnamefont{et~al.} (\bibinfo{collaboration}{WMAP}),
  \bibinfo{journal}{Astrophys. J. Suppl.} \textbf{\bibinfo{volume}{148}},
  \bibinfo{pages}{175} (\bibinfo{year}{2003}), \eprint{astro-ph/0302209}.

\bibitem[{\citenamefont{Babich et~al.}(2004)\citenamefont{Babich, Creminelli,
  and Zaldarriaga}}]{Babich:2004gb}
\bibinfo{author}{\bibfnamefont{D.}~\bibnamefont{Babich}},
  \bibinfo{author}{\bibfnamefont{P.}~\bibnamefont{Creminelli}},
  \bibnamefont{and}
  \bibinfo{author}{\bibfnamefont{M.}~\bibnamefont{Zaldarriaga}},
  \bibinfo{journal}{JCAP} \textbf{\bibinfo{volume}{0408}}, \bibinfo{pages}{009}
  (\bibinfo{year}{2004}), \eprint{astro-ph/0405356}.

\bibitem[{\citenamefont{Maldacena}(2003)}]{Maldacena:2002vr}
\bibinfo{author}{\bibfnamefont{J.~M.} \bibnamefont{Maldacena}},
  \bibinfo{journal}{JHEP} \textbf{\bibinfo{volume}{05}}, \bibinfo{pages}{013}
  (\bibinfo{year}{2003}), \eprint{astro-ph/0210603}.

\bibitem[{\citenamefont{Seery and Lidsey}(2005)}]{Seery:2005gb}
\bibinfo{author}{\bibfnamefont{D.}~\bibnamefont{Seery}} \bibnamefont{and}
  \bibinfo{author}{\bibfnamefont{J.~E.} \bibnamefont{Lidsey}},
  \bibinfo{journal}{JCAP} \textbf{\bibinfo{volume}{0509}}, \bibinfo{pages}{011}
  (\bibinfo{year}{2005}), \eprint{astro-ph/0506056}.

\bibitem[{\citenamefont{Vernizzi and Wands}(2006)}]{Vernizzi:2006ve}
\bibinfo{author}{\bibfnamefont{F.}~\bibnamefont{Vernizzi}} \bibnamefont{and}
  \bibinfo{author}{\bibfnamefont{D.}~\bibnamefont{Wands}},
  \bibinfo{journal}{JCAP} \textbf{\bibinfo{volume}{0605}}, \bibinfo{pages}{019}
  (\bibinfo{year}{2006}), \eprint{astro-ph/0603799}.

\bibitem[{\citenamefont{Lyth et~al.}(2003)\citenamefont{Lyth, Ungarelli, and
  Wands}}]{Lyth:2002my}
\bibinfo{author}{\bibfnamefont{D.~H.} \bibnamefont{Lyth}},
  \bibinfo{author}{\bibfnamefont{C.}~\bibnamefont{Ungarelli}},
  \bibnamefont{and} \bibinfo{author}{\bibfnamefont{D.}~\bibnamefont{Wands}},
  \bibinfo{journal}{Phys. Rev.} \textbf{\bibinfo{volume}{D67}},
  \bibinfo{pages}{023503} (\bibinfo{year}{2003}), \eprint{astro-ph/0208055}.

\bibitem[{\citenamefont{Sasaki et~al.}(2006)\citenamefont{Sasaki, Valiviita,
  and Wands}}]{Sasaki:2006kq}
\bibinfo{author}{\bibfnamefont{M.}~\bibnamefont{Sasaki}},
  \bibinfo{author}{\bibfnamefont{J.}~\bibnamefont{Valiviita}},
  \bibnamefont{and} \bibinfo{author}{\bibfnamefont{D.}~\bibnamefont{Wands}},
  \bibinfo{journal}{Phys. Rev.} \textbf{\bibinfo{volume}{D74}},
  \bibinfo{pages}{103003} (\bibinfo{year}{2006}), \eprint{astro-ph/0607627}.

\bibitem[{\citenamefont{Barnaby and Cline}(2007)}]{Barnaby:2007yb}
\bibinfo{author}{\bibfnamefont{N.}~\bibnamefont{Barnaby}} \bibnamefont{and}
  \bibinfo{author}{\bibfnamefont{J.~M.} \bibnamefont{Cline}},
  \bibinfo{journal}{JCAP} \textbf{\bibinfo{volume}{0707}}, \bibinfo{pages}{017}
  (\bibinfo{year}{2007}), \eprint{0704.3426}.

\bibitem[{\citenamefont{Arkani-Hamed et~al.}(2004)\citenamefont{Arkani-Hamed,
  Creminelli, Mukohyama, and Zaldarriaga}}]{ArkaniHamed:2003uz}
\bibinfo{author}{\bibfnamefont{N.}~\bibnamefont{Arkani-Hamed}},
  \bibinfo{author}{\bibfnamefont{P.}~\bibnamefont{Creminelli}},
  \bibinfo{author}{\bibfnamefont{S.}~\bibnamefont{Mukohyama}},
  \bibnamefont{and}
  \bibinfo{author}{\bibfnamefont{M.}~\bibnamefont{Zaldarriaga}},
  \bibinfo{journal}{JCAP} \textbf{\bibinfo{volume}{0404}}, \bibinfo{pages}{001}
  (\bibinfo{year}{2004}), \eprint{hep-th/0312100}.

\bibitem[{\citenamefont{Cheung et~al.}(2008)\citenamefont{Cheung, Creminelli,
  Fitzpatrick, Kaplan, and Senatore}}]{Cheung:2007st}
\bibinfo{author}{\bibfnamefont{C.}~\bibnamefont{Cheung}},
  \bibinfo{author}{\bibfnamefont{P.}~\bibnamefont{Creminelli}},
  \bibinfo{author}{\bibfnamefont{A.~L.} \bibnamefont{Fitzpatrick}},
  \bibinfo{author}{\bibfnamefont{J.}~\bibnamefont{Kaplan}}, \bibnamefont{and}
  \bibinfo{author}{\bibfnamefont{L.}~\bibnamefont{Senatore}},
  \bibinfo{journal}{JHEP} \textbf{\bibinfo{volume}{03}}, \bibinfo{pages}{014}
  (\bibinfo{year}{2008}), \eprint{0709.0293}.

\bibitem[{\citenamefont{Chen et~al.}(2007)\citenamefont{Chen, Easther, and
  Lim}}]{Chen:2006xjb}
\bibinfo{author}{\bibfnamefont{X.}~\bibnamefont{Chen}},
  \bibinfo{author}{\bibfnamefont{R.}~\bibnamefont{Easther}}, \bibnamefont{and}
  \bibinfo{author}{\bibfnamefont{E.~A.} \bibnamefont{Lim}},
  \bibinfo{journal}{JCAP} \textbf{\bibinfo{volume}{0706}}, \bibinfo{pages}{023}
  (\bibinfo{year}{2007}), \eprint{astro-ph/0611645}.

\bibitem[{\citenamefont{Chen et~al.}(2008)\citenamefont{Chen, Easther, and
  Lim}}]{Chen:2008wn}
\bibinfo{author}{\bibfnamefont{X.}~\bibnamefont{Chen}},
  \bibinfo{author}{\bibfnamefont{R.}~\bibnamefont{Easther}}, \bibnamefont{and}
  \bibinfo{author}{\bibfnamefont{E.~A.} \bibnamefont{Lim}}
  (\bibinfo{year}{2008}), \eprint{0801.3295}.

\bibitem[{\citenamefont{Komatsu and Spergel}(2001)}]{Komatsu:2001rj}
\bibinfo{author}{\bibfnamefont{E.}~\bibnamefont{Komatsu}} \bibnamefont{and}
  \bibinfo{author}{\bibfnamefont{D.~N.} \bibnamefont{Spergel}},
  \bibinfo{journal}{Phys. Rev.} \textbf{\bibinfo{volume}{D63}},
  \bibinfo{pages}{063002} (\bibinfo{year}{2001}), \eprint{astro-ph/0005036}.

\bibitem[{\citenamefont{Komatsu et~al.}(2005)\citenamefont{Komatsu, Spergel,
  and Wandelt}}]{Komatsu:2003iq}
\bibinfo{author}{\bibfnamefont{E.}~\bibnamefont{Komatsu}},
  \bibinfo{author}{\bibfnamefont{D.~N.} \bibnamefont{Spergel}},
  \bibnamefont{and} \bibinfo{author}{\bibfnamefont{B.~D.}
  \bibnamefont{Wandelt}}, \bibinfo{journal}{Astrophys. J.}
  \textbf{\bibinfo{volume}{634}}, \bibinfo{pages}{14} (\bibinfo{year}{2005}),
  \eprint{astro-ph/0305189}.

\bibitem[{\citenamefont{Komatsu et~al.}(2003)}]{Komatsu:2003fd}
\bibinfo{author}{\bibfnamefont{E.}~\bibnamefont{Komatsu}} \bibnamefont{et~al.}
  (\bibinfo{collaboration}{WMAP}), \bibinfo{journal}{Astrophys. J. Suppl.}
  \textbf{\bibinfo{volume}{148}}, \bibinfo{pages}{119} (\bibinfo{year}{2003}),
  \eprint{astro-ph/0302223}.

\bibitem[{\citenamefont{Babich}(2005)}]{Babich:2005en}
\bibinfo{author}{\bibfnamefont{D.}~\bibnamefont{Babich}},
  \bibinfo{journal}{Phys. Rev.} \textbf{\bibinfo{volume}{D72}},
  \bibinfo{pages}{043003} (\bibinfo{year}{2005}), \eprint{astro-ph/0503375}.

\bibitem[{\citenamefont{Creminelli et~al.}(2006)\citenamefont{Creminelli,
  Nicolis, Senatore, Tegmark, and Zaldarriaga}}]{Creminelli:2005hu}
\bibinfo{author}{\bibfnamefont{P.}~\bibnamefont{Creminelli}},
  \bibinfo{author}{\bibfnamefont{A.}~\bibnamefont{Nicolis}},
  \bibinfo{author}{\bibfnamefont{L.}~\bibnamefont{Senatore}},
  \bibinfo{author}{\bibfnamefont{M.}~\bibnamefont{Tegmark}}, \bibnamefont{and}
  \bibinfo{author}{\bibfnamefont{M.}~\bibnamefont{Zaldarriaga}},
  \bibinfo{journal}{JCAP} \textbf{\bibinfo{volume}{0605}}, \bibinfo{pages}{004}
  (\bibinfo{year}{2006}), \eprint{astro-ph/0509029}.

\bibitem[{\citenamefont{Smith and Zaldarriaga}(2006)}]{Smith:2006ud}
\bibinfo{author}{\bibfnamefont{K.~M.} \bibnamefont{Smith}} \bibnamefont{and}
  \bibinfo{author}{\bibfnamefont{M.}~\bibnamefont{Zaldarriaga}}
  (\bibinfo{year}{2006}), \eprint{astro-ph/0612571}.

\bibitem[{\citenamefont{Yadav et~al.}(2007)}]{Yadav:2007ny}
\bibinfo{author}{\bibfnamefont{A.~P.~S.} \bibnamefont{Yadav}}
  \bibnamefont{et~al.} (\bibinfo{year}{2007}), \eprint{0711.4933}.

\bibitem[{\citenamefont{Oh et~al.}(1999)\citenamefont{Oh, Spergel, and
  Hinshaw}}]{Oh:1998sr}
\bibinfo{author}{\bibfnamefont{S.~P.} \bibnamefont{Oh}},
  \bibinfo{author}{\bibfnamefont{D.~N.} \bibnamefont{Spergel}},
  \bibnamefont{and} \bibinfo{author}{\bibfnamefont{G.}~\bibnamefont{Hinshaw}},
  \bibinfo{journal}{Astrophys. J.} \textbf{\bibinfo{volume}{510}},
  \bibinfo{pages}{551} (\bibinfo{year}{1999}), \eprint{astro-ph/9805339}.

\bibitem[{\citenamefont{Smith et~al.}(2007)\citenamefont{Smith, Zahn, and
  Dore}}]{Smith:2007rg}
\bibinfo{author}{\bibfnamefont{K.~M.} \bibnamefont{Smith}},
  \bibinfo{author}{\bibfnamefont{O.}~\bibnamefont{Zahn}}, \bibnamefont{and}
  \bibinfo{author}{\bibfnamefont{O.}~\bibnamefont{Dore}},
  \bibinfo{journal}{Phys. Rev.} \textbf{\bibinfo{volume}{D76}},
  \bibinfo{pages}{043510} (\bibinfo{year}{2007}), \eprint{0705.3980}.

\bibitem[{\citenamefont{Bennett et~al.}(2003)}]{Bennett:2003bz}
\bibinfo{author}{\bibfnamefont{C.~L.} \bibnamefont{Bennett}}
  \bibnamefont{et~al.} (\bibinfo{collaboration}{WMAP}),
  \bibinfo{journal}{Astrophys. J. Suppl.} \textbf{\bibinfo{volume}{148}},
  \bibinfo{pages}{1} (\bibinfo{year}{2003}), \eprint{astro-ph/0302207}.

\bibitem[{\citenamefont{Serra and Cooray}(2008)}]{Serra:2008wc}
\bibinfo{author}{\bibfnamefont{P.}~\bibnamefont{Serra}} \bibnamefont{and}
  \bibinfo{author}{\bibfnamefont{A.}~\bibnamefont{Cooray}}
  (\bibinfo{year}{2008}), \eprint{0801.3276}.

\bibitem[{\citenamefont{Babich and Pierpaoli}(2008)}]{Babich:2008uw}
\bibinfo{author}{\bibfnamefont{D.}~\bibnamefont{Babich}} \bibnamefont{and}
  \bibinfo{author}{\bibfnamefont{E.}~\bibnamefont{Pierpaoli}}
  (\bibinfo{year}{2008}), \eprint{0803.1161}.

\bibitem[{\citenamefont{Cooray et~al.}(2008)\citenamefont{Cooray, Sarkar, and
  Serra}}]{Cooray:2008xz}
\bibinfo{author}{\bibfnamefont{A.}~\bibnamefont{Cooray}},
  \bibinfo{author}{\bibfnamefont{D.}~\bibnamefont{Sarkar}}, \bibnamefont{and}
  \bibinfo{author}{\bibfnamefont{P.}~\bibnamefont{Serra}}
  (\bibinfo{year}{2008}), \eprint{0803.4194}.

\bibitem[{\citenamefont{Jeong and Smoot}(2007)}]{Jeong:2007mx}
\bibinfo{author}{\bibfnamefont{E.}~\bibnamefont{Jeong}} \bibnamefont{and}
  \bibinfo{author}{\bibfnamefont{G.~F.} \bibnamefont{Smoot}}
  (\bibinfo{year}{2007}), \eprint{0710.2371}.

\bibitem[{\citenamefont{Taylor and Watts}(2001)}]{Taylor:2000hq}
\bibinfo{author}{\bibfnamefont{A.}~\bibnamefont{Taylor}} \bibnamefont{and}
  \bibinfo{author}{\bibfnamefont{P.}~\bibnamefont{Watts}},
  \bibinfo{journal}{Mon. Not. Roy. Astron. Soc.}
  \textbf{\bibinfo{volume}{328}}, \bibinfo{pages}{1027} (\bibinfo{year}{2001}),
  \eprint{astro-ph/0010014}.

\bibitem[{\citenamefont{Bernardeau and Kofman}(1995)}]{Bernardeau:1994aq}
\bibinfo{author}{\bibfnamefont{F.}~\bibnamefont{Bernardeau}} \bibnamefont{and}
  \bibinfo{author}{\bibfnamefont{L.}~\bibnamefont{Kofman}},
  \bibinfo{journal}{Astrophys. J.} \textbf{\bibinfo{volume}{443}},
  \bibinfo{pages}{479} (\bibinfo{year}{1995}), \eprint{astro-ph/9403028}.

\bibitem[{\citenamefont{Komatsu et~al.}(2008)}]{Komatsu:2008hk}
\bibinfo{author}{\bibfnamefont{E.}~\bibnamefont{Komatsu}} \bibnamefont{et~al.}
  (\bibinfo{collaboration}{WMAP}) (\bibinfo{year}{2008}), \eprint{0803.0547}.

\bibitem[{\citenamefont{Ma and Bertschinger}(1995)}]{Ma:1995ey}
\bibinfo{author}{\bibfnamefont{C.-P.} \bibnamefont{Ma}} \bibnamefont{and}
  \bibinfo{author}{\bibfnamefont{E.}~\bibnamefont{Bertschinger}},
  \bibinfo{journal}{Astrophys. J.} \textbf{\bibinfo{volume}{455}},
  \bibinfo{pages}{7} (\bibinfo{year}{1995}), \eprint{astro-ph/9506072}.

\end{thebibliography}

\end{document}